\documentclass[aps,prx, 
superscriptaddress, showpacs,twocolumn,nobalancelastpage,nofootinbib]{revtex4-1}

\usepackage{graphicx, color}
\usepackage{dcolumn}
\usepackage{bm}
\usepackage{amsmath}
\usepackage{amssymb}
\usepackage{verbatim}
\usepackage{appendix}
\usepackage{float}
\usepackage{color, soul}
\usepackage{gensymb}
\usepackage{enumitem}
\usepackage{xcolor}

\makeatletter
\renewcommand{\p@subsection}{}
\renewcommand{\p@subsubsection}{}
\renewcommand{\p@paragraph}{}
\makeatother

\usepackage{etoolbox}
\patchcmd{\section}
  {\centering}
  {\raggedright}
  {}
  {}
\patchcmd{\subsection}
  {\centering}
  {\raggedright}
  {}
  {}

\begin{document}


\title{Possible Applications of Dissolution Dynamic Nuclear Polarization in Conjunction with Zero- to Ultralow-Field Nuclear Magnetic Resonance}




\author{Danila A. Barskiy}
\email{dbarskiy@uni-mainz.de}
\affiliation{Institut f{\"u}r Physik, Johannes Gutenberg Universit{\"a}t Mainz, 55128 Mainz, Germany, \\
Helmholtz Institut Mainz, 55128 Mainz, Germany, \\
GSI Helmholtzzentrum f{\"u}r Schwerionenforschung, Darmstadt, Germany}

\author{John W. Blanchard}
\affiliation{Institut f{\"u}r Physik, Johannes Gutenberg Universit{\"a}t Mainz, 55128 Mainz, Germany, \\
Helmholtz Institut Mainz, 55128 Mainz, Germany, \\
GSI Helmholtzzentrum f{\"u}r Schwerionenforschung, Darmstadt, Germany}

\author{Dmitry Budker}
\affiliation{Institut f{\"u}r Physik, Johannes Gutenberg Universit{\"a}t Mainz, 55128 Mainz, Germany, \\
Helmholtz Institut Mainz, 55128 Mainz, Germany, \\
GSI Helmholtzzentrum f{\"u}r Schwerionenforschung, Darmstadt, Germany}
\affiliation{Department of Physics, University of California, Berkeley, CA 94720-7300 USA}

\author{Quentin Stern}
\affiliation{Univ Lyon, ENS Lyon, UCBL, CNRS, CRMN UMR 5082, F-69100, Villeurbanne, France}

\author{James Eills}
\affiliation{Institut f{\"u}r Physik, Johannes Gutenberg Universit{\"a}t Mainz, 55128 Mainz, Germany, \\
Helmholtz Institut Mainz, 55128 Mainz, Germany, \\
GSI Helmholtzzentrum f{\"u}r Schwerionenforschung, Darmstadt, Germany}

\author{Stuart J. Elliott}
\affiliation{Molecular Sciences Research Hub, Imperial College London, London W12 0BZ, United Kingdom}

\author{Rom\'an Picazo-Frutos}
\affiliation{Institut f{\"u}r Physik, Johannes Gutenberg Universit{\"a}t Mainz, 55128 Mainz, Germany, \\
Helmholtz Institut Mainz, 55128 Mainz, Germany, \\
GSI Helmholtzzentrum f{\"u}r Schwerionenforschung, Darmstadt, Germany}

\author{Antoine Garcon}
\affiliation{Helmholtz-Institut Mainz, 55099 Mainz, Germany}
\affiliation{Johannes Gutenberg-Universit{\"a}t Mainz, 55128 Mainz, Germany}

\author{Sami Jannin}
\affiliation{Univ Lyon, ENS Lyon, UCBL, CNRS, CRMN UMR 5082, F-69100, Villeurbanne, France}

\author{Igor V. Koptyug}
\affiliation{Laboratory of Magnetic Resonance Microimaging, International Tomography Center SB RAS, Institutskaya Street 3A, 630090 Novosibirsk, Russia}



\begin{abstract}
The combination of a powerful and broadly applicable nuclear hyperpolarization technique with emerging (near-)zero-field modalities offer novel opportunities in a broad range of 
nuclear magnetic resonance spectroscopy and imaging
applications, including biomedical diagnostics, monitoring catalytic reactions within metal reactors and many others. These are discussed along with a roadmap for future developments.
\end{abstract}


 \maketitle

\section{Introduction}

Zero- to ultralow-field nuclear magnetic resonance (ZULF NMR) has emerged from the far corner of NMR exotica and is becoming a relatively well established subfield of NMR on its own \cite{Blanchard2016emagres,Blanchard2021_LtL}. While for high-field NMR, hyperpolarization often represents a ``luxury'' of having enhanced signals, it is an absolute necessity for ZULF NMR where signals vanish in the absence of hyperpolarization. It is, therefore, not surprising that the progress in ZULF NMR techniques and applications has paralleled developments in hyperpolarization \cite{eills2023spin}. In fact, essentially all hyperpolarization techniques have by now been used with ZULF NMR \cite{Theis2011,Theis2012_NH_PHIP,Barskiy2019,Picazo2023,Chuchkova2023}.

At first glance, dissolution dynamic nuclear polarization (\textit{d}DNP) is not an obvious technique to use with ZULF NMR. In fact, $d$DNP usually involves relatively complex setup based on a high-field magnet and cryogenics, so it would seem to negate potential advantages of ZULF NMR such as portability, relative simplicity and low cost. Nevertheless, in this article\footnote{The paper is based on the discussions prior to February 2022 and projects initiated at that time.}, we argue that a  ``marriage'' of $d$DNP and ZULF NMR makes a lot of practical sense, given the growing availability of $d$DNP in preclinical and clinical contexts, the ability to hyperpolarize nearly arbitrary small molecules, and the compatibility of the two techniques with \emph{in vivo} studies due to their non-invasiveness \cite{JANNIN201941,Blanchard2021_LtL}. 

\subsection{What is ZULF NMR?}

ZULF-NMR experiments are performed at low bias magnetic fields, so that internal molecular interactions between spins are larger than interaction between the spins and the external field \cite{eMagRes-ZULF}. In most of the ZULF NMR apparatus today, detection is accomplished with spin-exchange relaxation-free (SERF) magnetometers \cite{ZULF-RSI}. At the heart of such a magnetometer is a glass cell containing atomic vapor, for example Rb, K, or Cs \cite{Budker2007}, heated to 150-180\,$^{\mathrm o}$C to ensure sufficiently high concentration of the alkali metal in the gas phase. Since the magnetic field from a polarized sample scales as $1 / r^3$ (where $r$ is a distance between the sample and the vapor cell), the vapor cell is typically positioned as close as possible to the sample.  Commercial magnetometers are currently available and the surface of such sensors remains relatively cool (40\,$^{\mathrm o}$C) on the outside \cite{Put2021}. 

The sensor and the sample are enclosed in a magnetic shield. The shield separates the ZULF region from the environmental magnetic fields (the Earth magnetic field, AC-line induced fields, etc.). External fields are typically attenuated by a factor of 10$^5$-10$^6$, while the residual fields are further reduced with a set of magnetic-field coils mounted inside the shield \cite{xu2006magnetic_PNAS}.

\subsection{Why $d$DNP?}

Dissolution dynamic nuclear polarization (\textit{d}DNP) is one of the key hyperpolarization methods used to enhance NMR signals of molecules in the solution-state  \cite{Larsen2003}. DNP works by applying a microwave field (with frequency set near-resonance with electron Zeeman transitions) to the solid-state sample kept at a high magnetic field containing molecules to be polarized and the source of electrons, typically, stable radicals like TEMPO \cite{VANDENBRANDT200453} or trityl \cite{SHANKARPALANI2023107411} (although other sources are known as well). The dissolution step involves fast injection of hot water and transfer of the hyperpolarized sample to a detection apparatus.

Over the last two decades, $d$DNP has emerged as a major player in various subfields of magnetic resonance, particularly, metabolic imaging using $^{13}$C-detected MRI (i.e., in the context of chemical shift imaging and single voxel spectroscopy) \cite{KURHANEWICZ201181}. It is therefore reasonable to explore opportunities provided by generality and widespread availability of $d$DNP for novel ZULF NMR/MRI applications.

Combining ZULF with $d$DNP makes even more sense in the current context, where there are ongoing efforts towards benchtop $d$DNP setups and with several teams working on making hyperpolarization generated by low-temperature off-site transportable over long distances \cite{ji2017transportable,el2021porous,capozzi2022design,steiner2023long}, with dissolution taking place remotely on-site, therefore possibly turning hyperpolarization in the near future into a consumable. The production of large quantities of hyperpolarized substrates on demand in a single site and the delivery of those hyperpolarized molecules on-site might end-up in overall lower costs and higher availability. 

\subsection{The purpose of this paper}

There have already been first reports of the combination of $d$DNP with ZULF NMR. In \cite{Barskiy2019}, a portable ZULF-NMR apparatus was taken to the University of California San Francisco (UCSF) hospital where a $d$DNP machine was available. Zero-field spectrum of $d$DNP-hyperpolarized [2-$^{13}$C]-pyruvic acid was recorded, demonstrating feasibility of the $d$DNP/ZULF-NMR combination.

Recent works \cite{Picazo2023,Mouloudakis2023} demonstrated ZULF-NMR detection of [$^{13}$C]-sodium formate, [1-$^{13}$C]-glycine, [2-$^{13}$C]-sodium acetate, and  [1–$^{13}$C]-pyruvate with \textit{d}DNP-provided signal enhancements of up to 11,000 compared to thermal prepolarization at 2\,T~\cite{ELLIOTT202159}. Relaxivity of 0.3\,s$^{-1}$mM$^{-1}$ was found for the TEMPOL radicals used in the $d$DNP process resulting in ZULF-NMR line broadening of 100\,mHz per mM of the radical. These results show that, despite the presence of the radicals, $d$DNP can be used as a universal hyperpolarization tool for ZULF-NMR applications. Further improvements will be possible in the near future to produce radical-free hyperpolarized solutions with heterogeneous sample formulations~\cite{Darai2021}.

In this paper, we discuss potential future applications of $d$DNP in conjunction with ZULF NMR, focusing on directions where such a combination may be particularly advantageous. 

\section{Applications of $d$DNP in ZULF NMR}

\subsection{Complex Environments and NMR of mixtures}\label{theme-complex-environments}

In conventional high-field NMR, the main source of chemical information is chemical shift wherein the applied magnetic field is partially shielded by molecular electrons, giving rise to a shift in the nuclear Larmor frequency. Because differences between chemical shifts corresponding to different chemicals are proportional to the applied magnetic field, high-resolution high-field NMR measurements require extremely homogeneous magnetic field and highly homogeneous samples. Any heterogeneity leads to susceptibility-induced magnetic field gradients that broaden nuclear spin resonance lines, reducing the resolution--and thus chemical specificity--of the NMR spectrum.

Unfortunately, many (if not most) important real-world materials and devices are inherently heterogeneous, including, for example, energy-storage devices and catalytic chemical reactors. While direct, non-destructive \textit{in situ} and/or operando measurements would be of tremendous value for understanding such systems, real samples and devices pose numerous challenges for traditional spectroscopic methods: in addition to the heterogeneity-based problems for NMR, many such devices are optically and RF-opaque and must remain fully sealed for practical safety considerations. 

Fortunately, a significant advantage of operating at zero magnetic field is that the resulting NMR spectra are almost entirely unaffected by the magnetic susceptibility of the sample environment. In fact, zero-field NMR experiments on heterogeneous samples routinely feature sub-Hz resonance linewidths \cite{Blanchard2015,Tayler2018}. Also, because low-field NMR involves frequencies much lower than in the high-field case, the skin effect becomes negligible (for example, the RF penetration depth of a 1\,kHz signal into stainless steel is $>1$\,cm). 

Apart from (or in addition to) the challenges of measurements in complex environments, another challenge is chemical analysis of complex mixtures with a representative example being the analysis of impurities in food and beverages.

\subsection{Wine and spirits testing} \label{Booze} 

The $d$DNP/ZULF-NMR spectroscopy can be used to screen for methanol content and other impurities in alcoholic beverages. This is of importance because
methanol poisoning is a major cause of injury and death in Eastern European and Asian countries \cite{giovanetti2013methanol,lee2014risk}; there is a general need to test for counterfeit and bootlegged products. Additional motivation may be that a similar method can be used for the detection of liquid poisons and explosives that escape nuclear-quadrupole-resonance (NQR) based screening.
High polarization available with $d$DNP is good for developing the technique and building up a library of relevant ZULF-NMR spectra.

ZULF NMR spectra are generally vastly different for different molecules,  with line widths much narrower than the major peak separation. This potentially facilitates analyte discrimination, even in low-cost instruments. 

\begin{figure}
    \centering
    \includegraphics[width=\columnwidth]{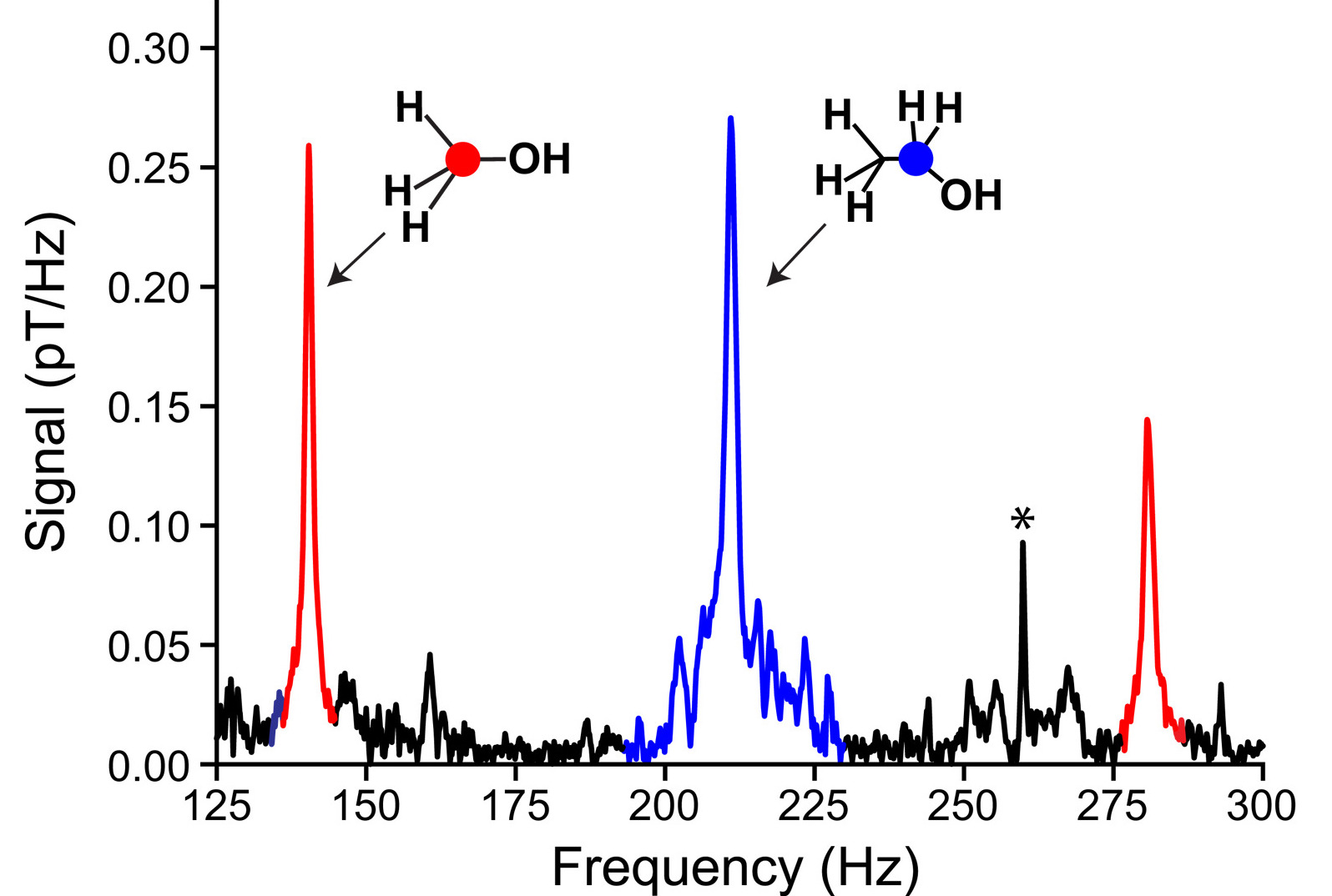}
    \caption{ZULF NMR spectrum of the sample 
    containing ethanol (750 mM) extracted from vodka and methanol (750 mM) at natural isotopic abundance with benzylamine (250 mM) and Ir catalyst (12 mM). The spectrum was observed following 1500 scans (each scan required 10 s of parahydrogen bubbling in a field of 19 mT at 5 bar and 60 sccm). The peaks of methanol at 140 and 280 Hz represent $J_{\rm CH}$ and $2J_{\rm CH}$, and the cluster of peaks surrounding 210 Hz arise from the AX$_2$ spin system of [1-$^{13}$C]-ethanol (1.5$J_{\rm CH}$, $J_{CH} = 140$ Hz). Peaks for [2-$^{13}$C]-ethanol are visible around 255 Hz, although markedly less clear than those of the other ethanol isotopomer; (*) denotes peak from an unknown source. Reproduced with permission from Ref.\,\cite{VanDyke2022}.}
    \label{fig:MeOH-EtOH}
\end{figure}

ZULF NMR spectra of ethanol extracted from a vodka sample (bought at a supermarket) was recently studied, where alcohol was hyperpolarized using parahydrogen \cite{VanDyke2022}. In particular, $J$-coupling transitions corresponding to [$^{13}$C]-methanol, [1-$^{13}$C]-ethanol and [2-$^{13}$C]-ethanol in both $^{13}$C-isotopically enriched and natural-abundance samples were observed after employing the relayed version of the signal amplification by reversible exchange (SABRE) hyperpolarization (Fig.\,\ref{fig:MeOH-EtOH}) \cite{Iali2018}. Interestingly, different lines in $J$-spectra showed different relaxation behavior, highlighting opportunities for a combined use of hyperpolarization and relaxometry for chemical identification. Fast Laplace-transform methods could be envisioned for obtaining 2D-spectra containing spectroscopic, relaxation, and diffusion information \cite{Telkki2018}.

Despite SABRE-relay being a convenient way of polarizing molecules containing exchangeable protons, it is still not as universal as $d$DNP as shown recently with the hyperpolarization of complex mixtures such as cancer or plant extracts to biological fluids \cite{dey2022fine, dumez2015hyperpolarized}. High levels ($>$5\%) of heteronuclear polarization using SABRE approach are yet to be demonstrated. Therefore, $d$DNP remains a primary method for universal hyperpolarization and ZULF NMR detection.

The spectrum of ${}^{13}$C-methanol has two peaks at 140 and 280\,Hz.
The spectrum of 1-${}^{13}$C-ethanol has peaks at near-zero frequency and around 200-220\,Hz.
The spectrum of 2-${}^{13}$C-ethanol has peaks at near-zero frequency, at around 120-130\,Hz, and around 230-260~Hz.
Overlap between the signals for ethanol and methanol should be small. However, the tails of the ethanol peaks may still be significant relative to the methanol signals if using a realistic samples (hopefully, with low methanol concentration).

When evaluating whether the $d$DNP/ZULF NMR analysis is the best way to go in practice, one also need to consider that what we have discussed so far is effectively a destructive analysis method, as the bottle has to be opened and the contents mixed with radicals and dissolution solvent. If, however, the analysis is a part of the production process, testing can be performed in real-time. One also needs to compare the $d$DNP/ZULF NMR against gas chromatography--mass spectrometry, another approach in which the sensor can be portable, and sensitive. 

We suggest further research on $d$DNP/ZULF NMR on ethanol and methanol which are good standard materials with well-understood spectra, which should lead to optimization of the analysis and extension to other possible impurities. 

\subsection{NMR of mixtures (liquids, adsorbed gases) in porous materials} \label{porous}
\vspace{-12pt}
An important task of NMR spectroscopy is to analyze (quantify) the composition of a mixture of fluids confined in a porous solid material.

There are numerous situations and processes which involve liquids or gases adsorbed/absorbed in porous materials, such as filtration of liquid mixtures, chromatographic separation, chemical (e.g., heterogeneous catalytic) reactions, transport of fluids in building materials and rock cores, etc. 
Analyzing the mixture composition in real time may be required as it may change in space and time because of adsorption, chemical transformation, etc. This can be potentially extended to adsorbed gases, but $d$DNP of gases \cite{Vuichoud2016,Comment2010} can be challenging as the relaxation rates are very much enhanced by spin rotation.

Hyperpolarization is generally required for such experiments because the porosity of solid materials is lower than unity, which reduces the amount of liquid in the sample under study, particularly for materials with low values of the specific pore volume. 
In this respect, $d$DNP is a most universal hyperpolarization technique at present, which can be used to polarize a broad range of fluids suitable for such studies.

\begin{figure}
    \centering
    \includegraphics[width=\columnwidth]{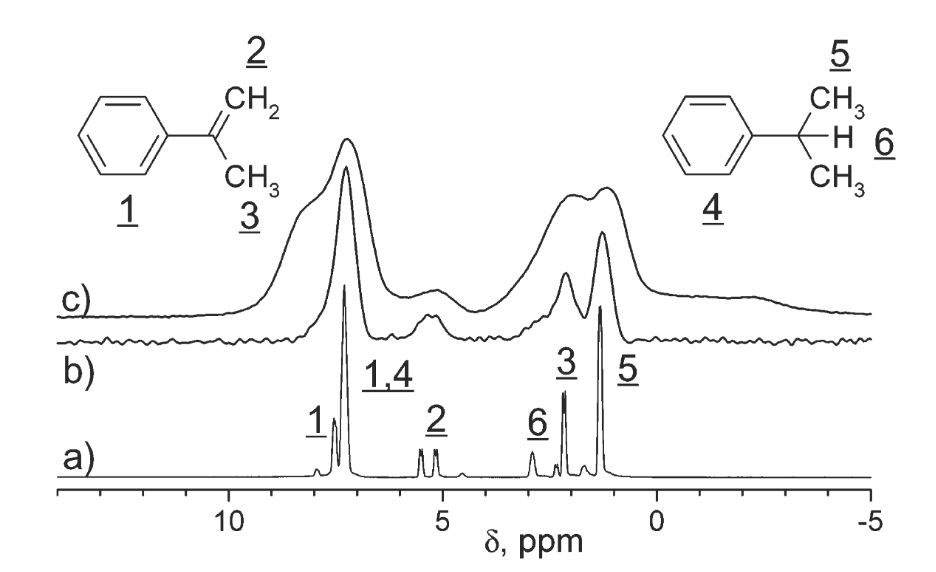}
    \caption{$^1$H  NMR spectra of a 1:1 mixture of $\alpha$-methylstyrene (AMS; left) and its hydrogenation product cumene detected for bulk liquid (a) and liquid permeating a porous catalyst pellet (b,c). Spectrum (b) was detected with spatial resolution. The number of acquisitions was eight (a,c) or two (b). Reproduced with permission from Ref.\,\cite{Koptyug2002}.}
    \label{fig:PorouHF}
\end{figure}

NMR spectra of fluids filling the pore space of porous solids are inevitably significantly broadened and distorted, with the main reason being the susceptibility mismatch between the involved liquid, solid, and gaseous (e.g., in partially saturated porous solids) phases which leads to large local gradients of the applied magnetic field when conventional NMR spectroscopy or imaging is performed (Fig.\,\ref{fig:PorouHF}).

\begin{figure}
    \centering
    \includegraphics[width=\columnwidth]{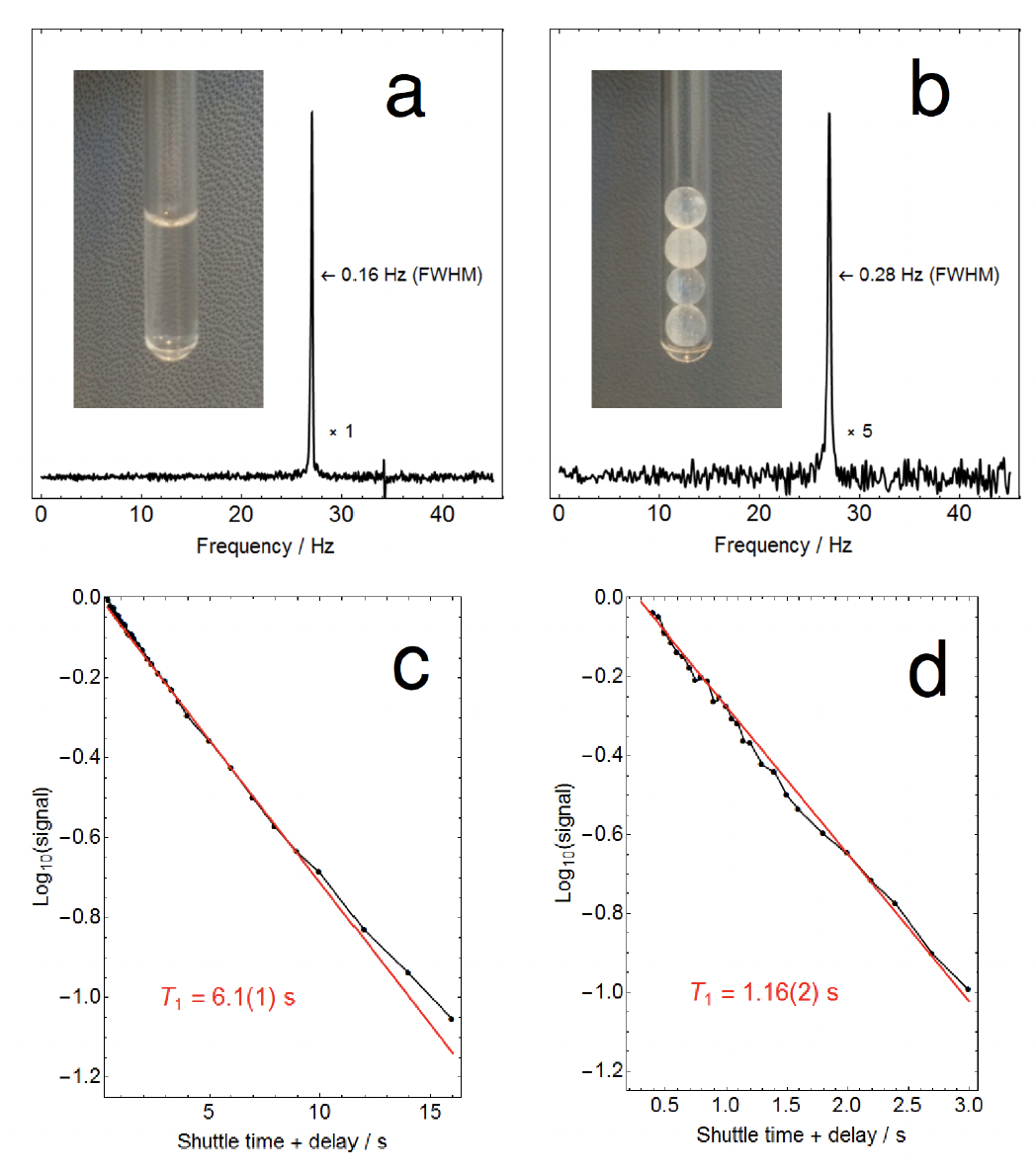}
    \caption{(a-b) Single-scan NMR spectra recorded using atomic magnetometer and (c-d) $T_1$ relaxation data recorded at a magnetic field of 0.634 $\mu$T (corresponding to $^1$H Larmor frequency of 27 Hz) for n-heptane in (a+c) bulk liquid and (b+d) imbibed into Q–15 porous silica. Reproduced with permission from Ref.\,\cite{Tayler2018}.}
    \label{fig:PorouZULF}
\end{figure}

Performing NMR experiments at lower magnetic fields (e.g., benchtop NMR systems) does not entirely solve the problem; while local field gradients become lower, the frequency separation of the NMR lines with different chemical shifts decreases as well, and thus no gain in spectral resolution is achieved this way. ZULF NMR, on the other hand, is advantageous in this respect since no (or very low) applied magnetic field means that no field gradients are present in the sample (Fig.\,\ref{fig:PorouZULF}), while at the same time chemical information is derived from the spin-spin couplings which are independent of magnetic field instead of chemical shifts.

\subsection{NMR of heterogeneous catalytic reactions} \label{catalysis}
Another promising direction related to the application of NMR spectroscopy is to monitor over time the conversion of substrates to products on porous solid catalysts. The general motivation for this is provided by the notion that NMR analysis of the reacting mixture is essential in characterizing the activity and selectivity of heterogeneous catalysts which are their most important characteristics from the point of view of industrial applications. In this context, hyperpolarization of nuclear spins \cite{eills2023spin} is essential for providing sufficient detection sensitivity for such experiments. In particular, as the concentrations of molecules vary in time during a catalytic reaction, high levels of initial hyperpolarization can help characterize the early (low concentration of products) and late stages of the reaction (low concentrations of a starting substrate). Furthermore, high levels of spin polarization may provide a unique possibility to detect short-lived reaction intermediates that are present in low concentrations during the reaction, which would be a tremendous asset for studying the underlying reaction mechanisms. In this respect, utilization of $d$DNP for spin hyperpolarization is essential as this technique currently outperforms other hyperpolarization approaches in terms of its much wider scope of potential reactants that can be hyperpolarized, and also often in terms of the achievable spin hyperpolarization levels. 

When it comes to the availability and quality of spectroscopic information, ZULF NMR is expected to have significant advantages over other NMR detection schemes because the application of any sizeable magnetic field results in its significant inhomogeneity for samples comprising both solid (catalysts and adsorbates) and fluid (reactants, products, solvent) phases. While such applications have not been demonstrated yet, their feasibility can be envisioned based on the successful application of $d$DNP to polarizing reactants to study homogeneous catalytic hydrogenation \cite{Boeg2019} and implementation of ZULF NMR to monitor the formation and consumption of hyperpolarized reaction products produced in sequential homogeneous hydrogenation of an alkyne with parahydrogen \cite{Burueva2020}.

\subsection{NMR spectroscopy and imaging of a model catalytic reactor}
Industrial (including catalytic) reactions often use high pressures and temperatures and thus reactors to implement such processes are often made of metals. Such reactors cannot be studied with conventional NMR/MRI. While it is possible to use different materials for constructing an NMR-compatible model reactor, this may not be suitable or realistic as non-traditional materials may affect reactor performance, for instance, because the heat transport in exothermic or endothermic reactions may be altered. Similar to the studies of heterogeneous catalytic processes addressed in the previous section, $d$DNP is necessary to provide adequate detection sensitivity, while ZULF NMR is required to achieve the necessary spectroscopic resolution. In addition, of primary importance in such studies is the fact that metal containers are not transparent to radiofrequency electromagnetic fields used for sample excitation and signal detection in conventional NMR/MRI because the penetration depth of rf field into metals is limited by the skin depth which is too small at conventional NMR frequencies (the MHz range). This problem is eliminated in ZULF NMR which does not use radiofrequency fields, opening the door for \emph{operando} reactor studies under harsh reaction conditions. Such spectroscopic (and potentially imaging) experiments are thus deemed feasible. Recent developments of fast transfer and injection devices for $d$DNP, based on liquid pumps operating at pressures up to 40 bars \cite{CEILLIER2021100017}, would readily provide  way to supply these reactors with hyperpolarized solutions even under high pressure conditions.

\subsection{Utilization and exploration of long-lived spin states} \label{LLS}
Another promising strategy is the production of long-lived spin states (LLSS) with $d$DNP and their observation with ZULF NMR. The utility of LLSS is in prolonging the useful time window for the observation of enhanced NMR signal \cite{LEVITT201969, Stevanato2015, Sonnefeld2023}, enabling obtaining NMR spectra with higher SNR and the study of slower physical and chemical processes than what is possible with hyperpolarized substances for which the lifetime of NMR signal enhancement is governed by the conventional $T_1$ nuclear spin-relaxation times. Dissolution DNP has already demonstrated the ability to efficiently produce LLSS of not totally symmetric molecules \cite{Vasos2009}, which is not universally possible with other hyperpolarization techniques. In addition, the preparation of LLSS with dissolution DNP ensures a major gain in SNR when such states are detected in NMR experiments. At the same time, many molecules that do not exhibit properties of LLSS in high magnetic fields may do so under ZULF NMR conditions because of the strong coupling of nuclear spins, and not only between nuclei of the same type but also between different nuclei \cite{Emondts2014}. At the same time, the correlated state of a spin system containing, for example, $^1$H and $^{13}$C nuclei, can be easily read out using ZULF NMR.

Relaxation of nuclear spin order at low magnetic fields may be affected by chemical exchange in the case of molecules containing exchangeable chemical groups, for example, protons (especially in conditions when $J$-coupling between the polarized nucleus and an exchanging group matches the chemical exchange rate). Studying relaxation dynamics and understanding which of the chemical pathways may lower the available hyperpolarization is therefore warranted. Since accelerated relaxation of chemically-exchanging systems is expected to occur at near-zero-field, it is reasonable to study these effects by ZULF NMR. Furthermore, the possibility to ``decouple'' the rapidly exchanging group from the target spin system by applying magnetic field pulses is attractive because it can in principle prolong the lifetime of hyperpolarization \cite{Alcicek2021}.

\subsection{Drug screening by $d$DNP-ZULF NMR} \label{drug-screening}
Pharmaceutical industry is in constant search for new methods to detect interactions between small molecules and biomolecules and NMR plays an important role in this field. 
Drug screening is an essential step of drug discovery that consists of testing a library of compounds as possible binders for a target protein (or another biomolecule). It requires reliable assays able to detect weak interactions between a small organic molecule and a large biomolecule, for which NMR is a technique of choice. 
However, the low sensitivity of NMR forces researchers to use large amounts of expensive proteins and work at high concentrations that sets a limit to the contrast between the bound and free state. $d$DNP has been proven as a means of overcoming these limitations \cite{KIM2019501} by allowing to decrease the ligand and protein concentrations.

In a $d$DNP-ZULF NMR experiment, one would hyperpolarize a small molecule (``the spy'') that is known to bind weakly to the protein and first detect it using ZULF NMR in the absence of protein. The experiment would be then repeated by injecting the hyperpolarized spy in a protein solution preloaded in the ZULF spectrometer. 
The relaxation (or, alternatively) coherence time of the spy is likely to be severely affected by binding (by changing rotational correlation time as well as by interactions with deuterons in the protein). 
Further experiments would be conducted adding competitor molecules to the protein solution. 
If the competitor binds to the protein, the hyperpolarized ligand will not bind and will thus (partially) recover the coherence time it has in the free state, which reveals the binding of the competitor. 
This type of ``competition experiment'' is common in drug screening. 

Detection at high field, although feasible, is not straightforward. 
In the case of experiments where ${}^1$H or ${}^{19}$F spins are detected, field inhomogeneity after injection (due to the uncertainty of injected volume produced by $d$DNP and microbubbles) causes large broadening. 
This limitation is expected to disappear at low field. 
Importantly, the relaxation contrast between bound and free form of a small molecule prepared in a LLSS is likely to be stronger at low fields \cite{Kowalewski2019book}. 

If the sensitivity is no longer the limitation, the concentrations can be set so as to have the optimal contrast. 
Among the possible proteins 
for such studies are heavy trans-membrane proteins that tend to aggregate so they cannot be kept in solution during the typical times required with standard thermal equilibrium NMR (typically 20 minutes). 
As a consequence, conventional NMR cannot always be used as it requires accumulation of scans.

\begin{figure}
    \centering
    \includegraphics[width=\columnwidth]{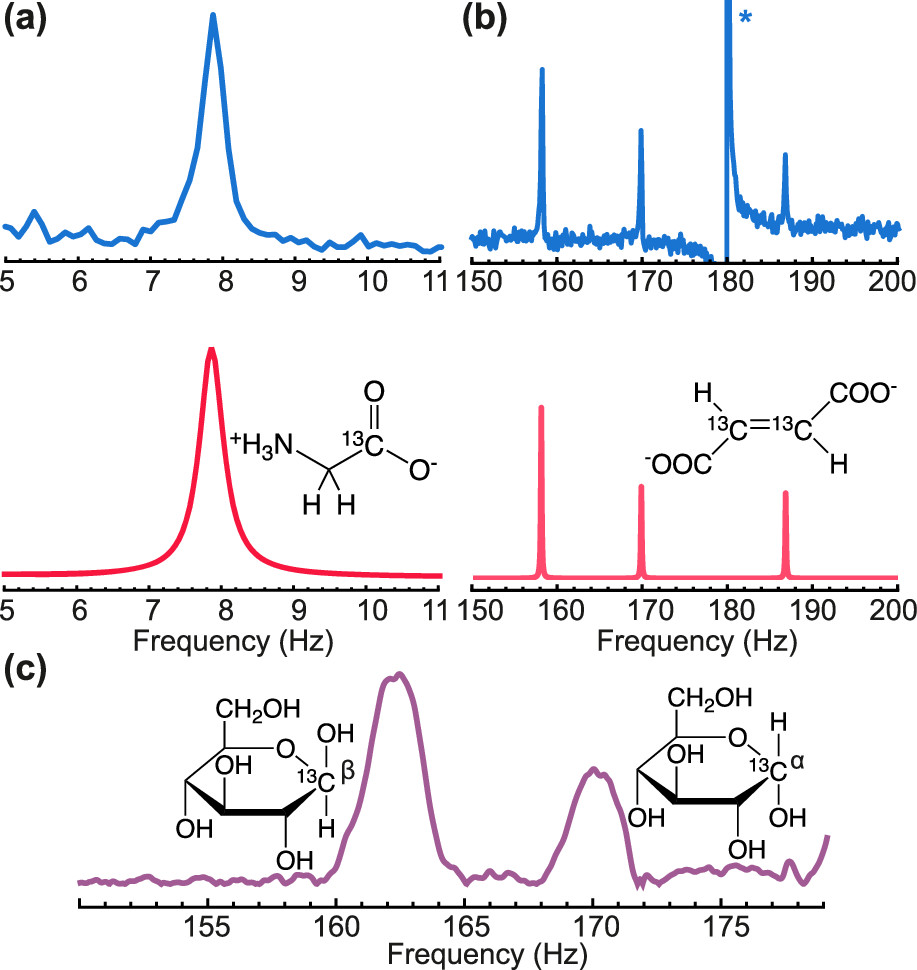}
    \caption{Zero-field $J$-spectra of (a) [1-$^{13}$C]-glycine, (b) [2,3-$^{13}$C$_2$]-fumarate, and (c) [1-$^{13}$C]-D-glucose. In (a,b), top pictures show experimental spectra after averaging (1024 scans) and the bottom spectra are simulations obtained in Mathematica using the SpinDynamica package. (25) In (c), the spectrum is a result of 2000 averages. The peak at 180 Hz denoted by asterisk is due to the residual magnetic field at the power-line harmonic leaking into the magnetic shielding. Reproduced with permission from Ref.\,\cite{Put2021}.}
    \label{fig:biomol}
\end{figure}

\subsection{Expanding the library of ZULF-NMR spectra}

For future practical applications, particularly, in the biomedical field, it will be useful to  collect high-quality ZULF-NMR spectra of a number of small biomolecules (Fig.\,\ref{fig:biomol}) \cite{Blanchard2020,Put2021}. In the first round, the focus could be on molecules/metabolites that are already under investigation for use as hyperpolarized sensors \textit{in vivo} for high-field MRI. The existing knowledge about these systems will help to evaluate the relative advantages and disadvantages of the ZULF approach. Another bonus is that some of these systems are already approved or are in the process of certification for use with patients.  One should note that strategies of accelerating DNP hyperpolarization buildup such as ${}^1$H-${}^{13}$C and ${}^1$H-${}^{15}$N cross-polarization enable satisfactory heteronuclear polarization only in a few tens of minutes.

Experimental data on ZULF NMR spectra are complementary to theoretical evaluations \cite{HOGBEN2011179} and serve to determine or validate the values of $J$-couplings \cite{Wilzewski2017} as well as various assumptions that enter the calculations of spin dynamics (e.g., the rates of relaxation and chemical exchange).

The acquisition of a substantially large number of spectra can be facilitated by modern automation techniques (robotic arms, microfluidics, remote control of the experiment, machine learning, etc.). 
Since it is easier to produce large homogeneous region of the low magnetic field (compared to high field), massive automated screening of different molecules could be employed at the same time \cite{Ashok2023Personal}.
Some of the important metabolites such as [1-${}^{13}$C]-pyruvate, [1-${}^{13}$C]-lactate, [1-${}^{13}$C]-fumarate, [1-${}^{13}$C]-malate, etc. have already been studied by ZULF NMR \cite{Barskiy2019,Put2021,eills2022metabolic,knecht2021rapid,theiss2023parahydrogen}. Other molecules, in particular,
${}^{15}$N-labelled compounds such as ${}^{15}$N-choline, ${}^{15}$N-phosphocholine, ${}^{15}$N-histidine, ${}^{15}$N-histamine, ${}^{15}$N-proline, ${}^{15}$N-trimethylglutamine among others, can be of interest to metabolomics, thus, their ZULF NMR investigation is warranted.
It is anticipated that the presence of low-gamma ${}^{15}$N nuclear sites may increase relaxation times of the observable transitions at ultralow fields, enabling a number of applications. One may need to pay careful attention to pH during polarization and/or detection to avoid exchange-related accelerated relaxation.

Until now, ZULF NMR has been mostly limited to either high-concentration solutions (often solvents), or parahydrogen-polarized molecules. With $d$DNP one have the possibility to collect ZULF spectra for many more classes of molecules by achieving polarization levels high enough to see them at biologically relevant concentrations. Besides adding to the library of ZULF NMR spectra, this is the first step towards eventual \emph{in vivo} imaging at zero field (see below).

\subsection{ZULF-NMR Molecular Imaging}

The goal of molecular imaging is to 
visualize, characterize, and quantify 
particular molecular species within intact subjects (such as materials, chemical reactors etc.) and organisms non-invasively and in real time \cite{gallagher2010introduction}.

Dissolution DNP is an enabling technique for \emph{in vivo} metabolic imaging \cite{KURHANEWICZ201181}, while ZULF NMR can be used to measure metabolites provided the signals are sufficiently enhanced by hyperpolarization \cite{Barskiy2019}. Thus, the combination of the two techniques appears to be both feasible and beneficial. Indeed, for metabolites present in complex environments all of the attractive features discussed in Sec.\,\ref{theme-complex-environments} are directly relevant.

Magnetic resonance imaging (MRI) is an important (if not the main) application of conventional NMR, and it is interesting to explore what can be done in the realm of ZULF. In  order to realize ZULF MRI, one needs to contend with the issue of concomitant gradients. According to Maxwell's equations, linear magnetic field gradients satisfy
\begin{equation}
    \frac{\partial B_z}{\partial z} = -\frac{\partial B_x}{\partial x} -\frac{\partial B_y}{\partial y}\,.
\end{equation}
While in high field, the influence of the transverse components (on the right-hand side) is typically negligible, this is not the case at ZULF conditions.

There are are several approaches currently being pursued to address the concomitant-gradient issue at ZULF, including:
\begin{enumerate}[label=(\roman*)]
\item \underline{Imaging with sensor arrays}.
	This approach is similar to magnetoencephalography (MEG), but offers additional spectroscopic information. While the inverse problem of reconstructing the source-magnetization distribution is ill-posed, information from a sensor array provides a rich input to modern analysis/reconstruction tools based on machine learning.
\item \underline{``Sweet-spot-scanning''}. Consider a quadrupolar field produced, for example, with an anti-Helmholtz coil pair. In this geometry, the field is zero at the center, and increases everywhere else. Thus, only a part of the sample sufficiently close to the center where the ZULF conditions are satisfied yield a coherent zero-field signal, while the rest of the sample would produce homogeneously broadened background. The ``sweet spot'' can be moved around to interrogate different parts of the sample and create its molecular image.
Because one would only be measuring a part of the total sample, it is imperative to start out with a lot of signal, which highlights the role of hyperpolarization in general and $d$DNP in particular. Note that more elaborate field configurations can be considered such as arrays of quadrupoles, which could allow simultaneous measurements at multiple field zeros to speed-up the imaging.

\item \underline{Fourier imaging}. Concomitant gradients are a problem at ultralow fields, but not so at higher fields. Correspondingly, spatial encoding can be achieved by applying a gradient on top of the field of the ``DC'' magnetic-field pulse used to initiate the ZULF-NMR signal. This approach resembles Hoult's ``Rotating-frame zeugmatography'' \cite{Hoult1979} or Transmit Array Spatial Encoding (TRASE) \cite{Sharp-TRASE}.
\end{enumerate}

\subsection{Dual-modality magnetoencephalography (MEG) and metabolic $J$-spectroscopy} \label{Metabolic-Brain-Imaging}
ZULF NMR spectra can be collected using same or similar devices as currently employed for MEG (see, for example, Ref.\,\cite{boto2018moving}). Detection of metabolic signatures associated with cognitive processes together with MEG could open new ways to better understand the human brain, for instance,
by detecting metabolic changes associated with brain activity. A combination of ZULF NMR with MEG for simultaneous mapping of the brain activity and brain chemistry could be used for detecting abnormalities and improving our general understanding of cognition.

Dissolution DNP is currently a leading hyperpolarization modality for clinical applications and, more likely, will remain to be so.
Optically pumped magnetometers (OPMs) are small, do not require cooling systems and can rival superconducting cryogenic quantum interference devices (SQUIDs) in terms of sensitivity \cite{boto2018moving}. At the same time, magnetic signals generated by hyperpolarized nuclei can be conveniently detected at zero field.

Let us estimate the sensitivity of OPM-based detection of $d$DNP-polarized molecules given the approximate parameters available with current setups. Let us assume injection of 40\,mL into a human patient of an aqueous buffer containing $C= 300$\,mM [1-$^{13}$C]-pyruvate with $^{13}$C polarization of $P = 40$\,\%. Assuming 5\,l of blood in a patient (and the fact that hyperpolarized bolus is redistributed within the blood reservoir in $\sim$20\,s \cite{UPTON1993}) and 100\,mL of blood at a given moment of time in the head, let us calculate $^{13}$C nuclear magnetization in a patient's head given hyperpolarized bolus is uniformly redistributed within the body without loss of hyperpolarization.

Magnetization ($M$) of such a bolus is
\begin{equation}
M = \frac{\gamma_{\rm C} \hbar}{2} C P \approx 6 \cdot 10^{-8} {\rm \frac{A}{m}}, 
\end{equation}
where $\gamma_{\rm C}$ is gyromagnetic ratio of $^{13}$C nucleus and $\hbar$ is Plank's constant. This magnetization generates a magnetic field measured with the OPM at the surface of the head (assuming magnetization is spread out inside the head uniformly, which is a crude approximation) at a distance $R=10$\,cm from the center of the head:
\begin{equation}
B = \frac{\mu_0 M}{2 \pi R^3} = 10 \, {\rm pT}, 
\end{equation}

This is plenty of signal since OPMs with $\sim$5 fT/Hz$^{1/2}$ sensitivity are now commercially available \cite{Blanchard2020}. In reality, relaxation during the transfer, injection, and distribution of the hyperpolarized material, as well as non-uniform distribution of polarization across the volume of the studied subject etc. will result in lower available signal. However, we point out that not $^{13}$C magnetization but rather heteronuclear spin coherence can be detected with ZULF NMR. Its signal is proportional to the difference between gyromagnetic ratios of the participating nuclei and the lifetime can exceed conventional $T_1$ relaxation of individual spins \cite{Emondts2014}, thus, even more signal can be expected than provided in the estimation above.

Another possible application is studying metabolic transformations in muscles during exercise/rest in order to reveal and quantify
metabolic changes associated with muscle activity. High-field MRI (with $d$DNP) can be used but it is limited to only a few muscle groups due to the small motion range restricted in conventional NMR scanners. We are aware of only one study that did it on calves muscles \cite{UPTON1993}. Future zero-field NMR wearable spectrometers will be ideally suited for such studies. 

\subsection{Production and Detection of Nuclear Spin Isomers}\label{Spin-Isomers}
Nuclear spin isomers (NSIs) have been explored in the context of fundamental spin physics, chemistry \cite{chapovsky1999nuclear}, and even astronomy \cite{curl1966spin}. Here we suggest producing NSIs by means of $d$DNP and measuring their lifetimes with ZULF NMR.

Currently, there is no way to obtain large quantities of NSIs for molecules other than hydrogen (for the latter, the well-known spin isomer, parahydrogen, is routinely produced \cite{Farkas1935}). Current techniques for NSI enrichment employ laser cooling \cite{chapovsky1999nuclear} and Stern-Gerlach spin-separation devices \cite{wennerstrom2012stern}. These methods cannot yield sufficiently large quantities of NSIs since only several milliliters can be stored in a flask or a tank, while larger quantities are desirable for both research and applications. 

$d$DNP is universal hyperpolarization technique and allows high ($>$50\%) polarization levels. SABRE can be used for polarizing NH$_3$ \cite{Iali2018} but sufficiently high and reproducible polarization levels ($>5$~\%) have not been demonstrated as of yet.
The advantage of ZULF NMR for NSI detection is that the spectra will directly show the distribution of NSIs via the intensity of $J$-spectral lines (a chemical step inducing molecular symmetry breaking would be necessary in some cases). Another advantage of ZULF NMR is the fact that modular spectrometers can be easily optimized for various chemical manipulations (which can be necessary in some cases to observe spin dynamics) Simultaneous ZULF detection of NSI with IR/Raman spectroscopy could also be envisioned.

\subsection{Developing Decoupling Sequences for ZULF NMR} \label{Is it useful?}
One of the unique advantages of ZULF NMR is that the technique is, in principle, capable of detecting spin-spin couplings that are ``truncated'' in high-field NMR. However, practical realization of this advantage requires the ability to suppress larger ``background'' interactions, which can be accomplished with rank-selective pulse sequences to decouple terms in the Hamiltonian of different rank. An example is decoupling the rank-2 dipole-dipole interactions so that they do not overwhelm the antisymmetric rank-1 $J$ coupling in an oriented sample \cite{Llor1995a,Llor1995b,King2017,Blanchard2020PNC}.

\subsection{Searches  for dark matter and exotic spin-gravity coupling} \label{CASPEr}

\begin{figure}
    \centering
    \includegraphics[width=\columnwidth]{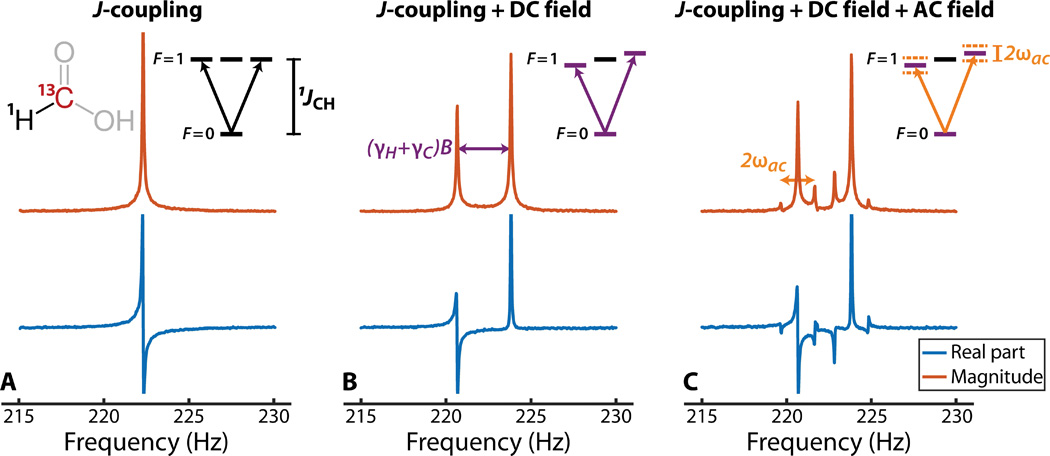}
    \caption{Nuclear spin energy levels and NMR spectra of $^{13}$C-formic acid measured in three different field conditions. (A)~At zero magnetic field, the $F = 1$ levels are degenerate, resulting in a spectrum exhibiting a single peak at the $J$-coupling frequency. (B)~In the presence of a DC magnetic field $B_z \approx 50$~nT, the $m_F = \pm 1$ degeneracy is lifted. The spectrum exhibits two split $J$-resonances. The splitting is equal to $\hbar B_z ( \gamma_{\rm C} + \gamma_{\rm H} )$. The asymmetry of the resonances is due to the influence of the applied field on the response characteristics of the atomic magnetometer. (C) Addition of an oscillating magnetic field along $B_z$ modulates the $m_F = \pm 1$ energy levels, resulting in sidebands located at $ J/2 \pi \pm B_z ( \gamma_C + \gamma_H ) / 2 \pi \pm \omega_{AC}$ with amplitude proportional to the modulation index: $A_s \propto B_{\rm AC} ( \gamma_{\rm C} + \gamma_{\rm H} ) / ( 2 \omega_{\rm AC} )$. Reproduced with permission from Ref.\,\cite{Garcon2019}}
    \label{fig:enter-label}
\end{figure}

ZULF NMR has been proven to be a powerful technique with potential to detect ultralight bosonic fields hypothesized to be constituents of galactic dark matter responsible for about 80\% of galactic mass that appears to be ``invisible'' apart from its gravitational pull. This was accomplished in the cosmic-axion spin-precession experiments (CASPEr) \cite{Wu2019Search,Garcon2019} that searched 
for axion-like particles (ALPs) and dark-photon (DP) dark-matter particles.
If those particles were to exist, they should induce phase and frequency modulation/shifts of the well-defined ZULF-NMR spectra of small molecules through their interaction with nuclear spins. However, the interaction between ALPs and nuclear spins is predicted to be feeble and hyperpolarization of nuclear spins needs to be employed to enable competitive sensitivity. We note that CASPEr-ZULF experiments take advantage of the ``intramolecular comagnetometry'' to distinguish the sought-after effect from systematics \cite{Ledbetter2012liquid,Wu2018}.

We propose to utilize $d$DNP techniques to polarize nuclear spins prior to the acquisition, potentially improving the sensitivity by five orders of magnitude compared to the experiments proposed in Refs. \cite{Wu2019Search,Garcon2019}, thus, enhancing the dark matter detection potential. Detection of dark matter via a non-gravitational interaction would resolve one of the biggest mysteries of modern physics by uncovering the nature this phenomenon. 

Hyperpolarization is necessary to perform dark-matter NMR searches, as the signals we aim to observe are extremely weak. $d$DNP is currently the most general method, enabling hyperpolarization on the widest range of spin targets.
ZULF NMR atomic magnetometers are among the most sensitive sensors for signals in the 0-500\,Hz frequency range. This places ZULF NMR as promising method to search for ALPs and DPs within this range for which only a few experiments exists even though the existence of ALPs and DPs in this range is well motivated (see, for example, \cite{wei2023dark} and references therein).

Another important fundamental-physics application of ZULF NMR with hyperpolarized samples is the search for a possible coupling of nuclear spins to gravity (the Hamiltonian term proportional to $\vec{S}\cdot \vec{g}$ with $\vec{S}$ being the spin and $\vec{g}$--the local acceleration due to gravity) \cite{Ledbetter2012liquid,Kimball2017SpinGravity,Wu2018}. Such an interaction would violate several fundamental symmetries; however, it is hypothesized in certain theories of gravity and represents an important probe into the connection between gravity and quantum mechanics \cite{Kimball2023,Zhang2023}.

Consideration for practical realization of $d$DNP-ZULF-NMR fundamental physics experiments include ensuring repeatable and stable supply of hyperpolarized material with continuous (automated) data acquisition to enable multi-shot averaging, calibration, and auxiliary experiments to identify and control for systematic errors.

\section{Conclusions and outlook}

The a-priori non-obvious combination of $d$DNP hyperpolarization with ZULF NMR can, indeed, be advantageous in a broad range of practical applications including but certainly not limited to the ones discussed above. The particularly powerful aspects of this combination include the ability to hyperpolarize a broad range of samples, insensitivity to sample inhomogeneities, and the ability to conduct spectroscopic measurements and imaging within conducting (e.g., metal) containers. In the medical context, one can benefit from the availability of $d$DNP infrastructure at hospitals \cite{ARDENKJAERLARSEN20163}, the relative simplicity, compactness and portability of ZULF-NMR devices \cite{Blanchard2020,Put2021}. These features of ZULF NMR are likely to be augmented in the near future when more compact magnetometers based on color centers in diamond \cite{Zheng2019ZeroField} come into use. It is our hope that this overview will serve as a ``roadmap'' to stimulate developments in this area.

\section{Acknowledgements}

We thank Prof.\,Blümich for his interest in our work and helpful advice he has given us over the years. ZULF-NMR imaging was extensively discussed with Dr. Nataniel Figueroa Leigh. 

\section{Declarations}
 
\subsection{Ethical Approval}
Not applicable.
 
\subsection{Competing interests}
The authors declare that they have no conflict of interest.
 
\subsection{Authors' contributions}
All authors contributed ideas. DAB and DB wrote the final text. All authors reviewed the manuscript.

\subsection{Funding}
This work was supported in part by the DFG/ANR grant BU 3035/24-1; DAB acknowledges financial support from the Alexander von Humboldt Foundation in the framework of Sofja Kovalevskaja Award; QS, SJE, and SJ thank the ENS-Lyon, the French CNRS, Lyon 1 University, the European Research Council under the European Union’s Horizon 2020 research and innovation program (ERC Grant Agreements No. 714519/HP4all and Marie Skłodowska-Curie Grant Agreement No. 766402/ZULF), and Bruker Biospin; IVK thanks the Russian Science Foundation (grant \#22-43-04426). 
 
\subsection{Availability of data and materials}
This manuscript is a roadmap paper as such it reviews the current status of the work in this area and proposes future directions.

\bibliography{d-DNP+ZULF}

\end{document}